Imaging the evolution of metallic states in a spin-orbit interaction driven correlated iridate

Yoshinori Okada[1], Daniel Walkup[1], Hsin Lin[2], Chetan Dhital[1], Tay-Rong Chang[3], Sovit Khadka[1], Wenwen Zhou[1], Horng-Tay Jeng[3], Arun Bansil[2], Ziqiang Wang[1], Stephen Wilson[1], Vidya Madhavan[1]

The Ruddlesden-Popper (RP) series of iridates ($Sr_{n+1}Ir_nO_{3n+1}$) have been the subject of much recent attention due to the anticipation of emergent physics arising from the cooperative action of spin-orbit (SO) driven band splitting and Coulomb interactions[1-3]. However an ongoing debate over the role of correlations in the formation of the charge gap and a lack of understanding of the effects of doping on the low energy electronic structure have hindered experimental progress in realizing many of the predicted states[4-8] including possible high-$T_c$ superconductivity[7,9]. Using scanning tunneling spectroscopy we map out the spatially resolved density of states in the *n*=2 RP member, $Sr_3Ir_2O_7$ (Ir327). We show that the Ir327 parent compound, argued to exist only as a weakly correlated band insulator in fact possesses a substantial ~130meV charge excitation gap driven by an interplay between structure, SO coupling and correlations. A critical component in distinguishing the intrinsic electronic character within the inhomogeneous textured electronic structure is our identification of the signature of missing apical oxygen defects, which play a critical role in many of the layered oxides. Our measurements combined with insights from calculations reveal how apical oxygen vacancies transfer spectral weight from higher energies to the gap energies thereby revealing a path toward obtaining metallic electronic states from the parent-insulating states in the iridates.

[1]Department of Physics, Boston College, Chestnut Hill, Massachusetts 02467, USA

[2]Physics Department, Northeastern University, Boston, Massachusetts 02115, USA

[3]National Tsing Hua U., Hsinchu, Taiwan and Institute of Physics, Academia Sinica, Taipei, Taiwan

Spin orbit (SO) coupled materials are at the forefront of the ongoing search for fundamentally new electronic states. Novel emergent phases such as the helical Dirac states of topological insulators[10,11] and exotic magnetic states[12] such as Skyrmion lattices[13] are recent direct manifestations of SO coupling. Most of these materials however are well described by single particle physics, with correlations playing a negligible role in determining their properties. A new frontier in this search, with many predictions of as yet undiscovered phases, is the exploration of materials where SO coupling and correlations are both relevant. The 5d-transition metal iridium oxides (iridates) have been proposed as excellent candidates for materials in this category[1-3].

Correlations in the 5d and 4d transition metal oxides are expected to be weaker than in their 3d counterparts. Following this expectation, many of 3d transition metal oxides show Mott insulating ground states[14], while the 4d-ruthenates show mostly metallic behavior. Surprisingly, despite the anticipation of even weaker correlation effects in the 5d compounds, many of iridates display magnetic, non-metallic properties[1-3]. The concept proposed to explain the origin of the insulating behavior of 5d-iridates is that the stronger SO interaction in the iridates creates additional band splitting, resulting in narrow bands close to the Fermi energy, thereby enhancing correlation effects[1-3]. In fact, when correlations are strong enough, a novel SO interaction driven $J_{eff}$=1/2 Mott insulator (spin-orbit Mott insulator) is predicted to emerge. This idea has led to theoretical proposals of exciting emergent phenomena including the potential appearance of high-Tc superconductivity[7]. So far however, the validity and universality of this concept in the 5d-iridates has been under intense debate[15-18].

Amongst the iridate families, the RP series ($Sr_{n+1}Ir_nO_{3n+1}$) goes through a transition from an insulator to metal with increasing *n*. Within this series the *n*=2 compound $Sr_3Ir_2O_7$ (Ir327) occupies a unique place, straddling a well-defined insulator (*n*=1) on one side and a metal (*n*=∞) on the other[3,16]. This places Ir327 in close proximity to a delicate transition point where the interplay between the comparable energy scales of SO and on-site Coulomb interactions (*U*) can produce novel effects within the charge and spin degrees of freedom. Experimental results on Ir327[3,18,19] including transport, diffraction, and optical spectroscopy demonstrate multiple correlated order parameters and phase transitions indicative of a ground state paralleling those found in their strongly correlated 3d transition metal counterparts. However, the role of correlations remains controversial, with recent suggestions that the material is a simple band insulator with only weak residual correlations[9]. Furthermore, contradictory experimental reports have made the resolution of these issues even more difficult. For example, optical spectroscopy measurements reveal a negligible gap in Ir327[3], while recent angle resolved photoemission spectroscopy (ARPES) data show indications of a substantially larger gap[20,21]. The inherent complexities arising from the fact that Ir327 is potentially close to a Mott transition including the tendency for electronic inhomogeneity often seen in such systems[22,23] have created experimental challenges resulting in important unanswered questions on the actual gap size, the role of correlations, and the influence of defects and dopants on the low energy density of states.

Spatially resolved density of states measurements across the Fermi energy, which can provide key information for resolving these issues have not yet been performed on the iridates. In this work, we measure high quality single crystals of Ir327 with scanning tunneling microscopy (STM) and spectroscopy. Our local density of states maps *(LDOS(r,eV)* =

*dI/dV(r,eV))* reveal an inhomogeneous, textured electronic landscape. Within the spatially varying density of states (DOS), we identify the intrinsic DOS of Ir327 with a hard gap of ~130meV. Comparisons with GGA calculations clearly show that this gap arises from a combination of rotated oxygen octahedra, enhanced spin-orbit coupling and coulomb interactions with each of these playing critical roles in determining the low energy density of states. The data further reveal that the metallic regions are correlated with single atom defects. Remarkably our *dI/dV(r,eV)* maps allow us to identify these defects as apical oxygen vacancies via a direct visualization of the rotated oxygen octahedra in real-space. This offers us a unique opportunity to determine how apical oxygen vacancies alter the local DOS from insulating to metallic. These data provide critical information for tuning the low energy electronic structure of iridates; an important component for realizing novel emergent phases such as superconductivity. Equally importantly, our measurements open the doors to STM investigations of this important family of iridates.

Single crystals of $Sr_3Ir_2O_7$ were grown via flux techniques similar to earlier reports[19]. The Ir327 single crystals thus prepared were cleaved at ~77K in ultra high vacuum before being directly transferred to the STM head held at 4K. Figure 1 shows STM topographic images of the surface. Similar surfaces were reproducibly imaged on multiple samples by different tips as long as the crystals were cleaved at low temperatures. The natural cleaving plane is in between the perovskite bilayer structure, whose unit length along c-axis corresponds to half of lattice constant *c* (Fig. 1a and b). Cleaving is expected to result in the exposure of the strontium oxide (SrO) plane (Fig. 1a). The topography (Fig. 1c and d) and its Fourier transforms (FT) (Fig. 1e) show a 1X1 square lattice of atoms separated by 3.9 Å, which is a distance equal to the separation between either strontium or apical oxygen atoms. The FT of the topography shows additional peaks at the √2x√2 positions (Fig. 1e). Similar features in the isostructural ruthenate $Sr_3Ru_2O_7$, were attributed to two possible causes, either the broken symmetry arising from alternating rotations of the oxygen octahedra surrounding the Ru atoms[24], which also exists for our Ir327 samples, or a charge density wave (CDW) formation on the surface[25]. In our case, since √2x√2 structure persists to high energies, it cannot be attributed to a simple CDW. While the actual origin of the √2x√2 structure is not completely clear, since this topographic feature has no bearing on the main points of current study, we do not pursue it any further in this paper. The topographic images show two types of chemical disorder; the bright dots and the square-shaped patterns indicated by arrows in Fig. 1c and circles in Fig. 1d, respectively. The bright dots are most likely adsorbed impurities on top of the surface and can be attributed to either excess Sr or oxygen atoms probably created during cleaving process. On the other hand, the defects shown in Fig. 1f-i (square-shaped defects with two different chiralities) arise from embedded surface or sub-surface defects whose identity will be discussed in greater detail later in this paper.

While the topographies at negative voltages (Fig. 1c) are reasonably homogenous, those at positive voltages (Fig. 1d) show bright and dark patches indicating a highly inhomogeneous electronic structure with a characteristic length scale of a few nanometers. To understand the origins of this inhomogeneity and the behavior of the density of states (DOS), we plot the tunneling spectra obtained along a line cut across a chiral defect as shown in Fig. 2a. We find that the LDOS exhibit dramatic changes across defects, evolving from spectra showing almost zero LDOS near $E_F$ (spectra close to bottom in Fig. 2c and e) to more a metallic V-shape (spectra close to the top in Fig. 2). This poses the question: what is the intrinsic DOS of the parent compound of Ir327 and what is responsible for this inhomogeneity? In semiconductors and insulators with poor screening, random dopant distribution is often responsible for electronic inhomogeneity. In our case, although the material is not deliberately doped, there are naturally occurring defects that may be potentially responsible for this behavior. Comparing the STM topography (Fig. 2a) to the spectral line cut (Fig. 2b-e), it is clear that the V-shaped LDOS is correlated with the presence of the square shaped chiral defect. In contrast, the spectra away from defects, display a highly suppressed (approximately zero) LDOS from ~10 meV below the $E_F$ to ~120 meV above $E_F$ (Fig. 2b and c). Zooming out to a wider energy range (Fig. 2d and e) we find that these regions show a semiconducting/insulating line shape with the $E_F$ positioned at very close to a band edge. This indicates that the intrinsic DOS in the ground state of parent Ir327 system is a ~130 meV hard gapped insulator.

Having determined the intrinsic DOS of parent Ir327, we address the role and importance of correlations in determining the electronic structure of this system. GGA calculations of the band structure (Fig. 3a-c) and the DOS for various values of $U$ (Fig. 3d) are shown in Fig.3. As per our calculations, increasing both SO (which lowers the occupied band at $\Gamma$ as shown in Fig. 3b) and onsite $U$ (which raises the unoccupied band at the $\Sigma$ point, as shown in Fig. 3c) are needed to explain our experimentally observed gap in the LDOS near the Fermi energy (also see Supplementary Information Fig.S2 for more detail). We note that the calculations were performed within a larger parameter space by varying the SO coupling as well as the octahedral in-plane tilt angle (within realistic deviations of a few degrees from the measured values), and these parameters (including the enhanced SO coupling of 1.7) were chosen to match the recent ARPES reports of band dispersion[20,21]. Although GGA calculations are not expected to capture the quantitative magnitude of the gap, the failure of the $U$=0 calculation to open any charge excitation gap at all indicates that the crystal field effects combined with strong spin orbit coupling are grossly insufficient to create gap of the magnitude observed by us. The conclusion of our studies then is that correlations are critical to understanding the band structure of the parent iridate Ir327. The natural question then is how does this correlated band structure evolve into the more metallic density of states?

To answer this question we focus on the V-shaped LDOS related to the chiral defects. Our first task is to determine their chemical identity. As shown in Fig. 1f-i, these defects retain $C_4$ symmetry while breaking reflection symmetry locally, and are aligned along the high symmetry directions of the lattice seen in the STM topographic images (green dots in Fig. 1f-i). One can easily imagine that the different chiralities (Fig. 1f and g) indirectly reflect the in-plane octahedral rotations that lead to two inequivalent lattice sites. In our case, the validity of this conjecture can be directly visualized by analyzing the LDOS maps. As shown in Fig. 4a, positive energy LDOS maps reveal square shaped patterns associated with the chiral impurities. Closer examination shows that unlike their shapes in the topography, the LDOS squares are not aligned with the lattice and are in fact distinctly rotated about the c-axis in either a clock wise (CW) or counter clock wise (CCW) manner with respect to the a-axis as shown in Fig. 4b and c. Plotting the histogram of rotation angles as measured from the a-axis (Fig. 4d), we find a bimodal distribution of $(12.5 \pm 2)°$. The bimodal nature of the rotations and the closeness of this absolute value to the measured in-plane octahedral rotation angle (~12°) from x-ray diffraction[26] indicate that our LDOS maps capture the local symmetry of the iridium oxide planes (Fig. 4e-g). Based on the experimentally observed rotational symmetry of the defects, we rule out the possibility that these impurities are centered at Sr sites (Fig. 4f), which do not reflect the ~12° octahedral rotation. Instead we identify the defect center with vacancies in either the apical oxygen site in the SrO plane or Ir site in the IrO$_2$ plane (Fig. 4g). Furthermore, the fact that oxygen vacancies are common defects in layered oxides is a strong indicator that the chiral defects are vacancies at the apical oxygen sites. We note here that one of the most difficult tasks in an STM experiment is to identify the chemical identity of the imaged atoms. However, as shown here, utilizing LDOS patterns reflecting the symmetry of the underlying lattice provides an as yet unexplored method of determining the identity of imaged atoms. In this case for example, the identification of the impurity center site within the lattice enables us to determine that the imaged atoms in the topography must be Sr atoms, laying to rest the controversy of which atom is imaged in STM topographies of SrO surfaces in the ruthenates and iridates.

Having identified the defect species with apical oxygen vacancies, we are now ready to discuss the effects of these vacancies on the low energy density of states. In many oxides including the cuprates, apical oxygen sites play a critical role in determining the functionality of the system[27-29]. Our ability to locate and measure the density of states at apical oxygen vacancy sites provides us with a unique opportunity to determine their effects as a function of both position and energy. To do this systematically, we analyze the LDOS in the entire region shown in Fig. 1c and d. First, by comparing the topographic images (Fig. 1c and d) and the LDOS maps (Fig. 5a-e), we see that the higher DOS regions (red) are correlated with the apical oxygen vacancy locations. The spatial distribution of the DOS thus suggests a method to obtain a systematic picture of the effects of these vacancies on the DOS by categorizing the corresponding spectra based on conductance values. We first classify all the spectra obtained in this region into 20 groups based on the conductance at +100 mV (also see Supplementary Information for more details) and then average the spectra within each category. The averaged spectra are shown in Fig. 5f and g. Here, $N$=1 represents the index for the highest and therefore most metallic LDOS (top red spectra in Fig. 5f and g) and $N$=20 represents the index for close to zero i.e., insulating LDOS (bottom purple spectra in Fig. 5f and g). By comparing the topographic images (Fig. 1c and d), LDOS maps (Fig. 5a-e), and the LDOS spectral shapes (Fig. 5f and g), we confirm that areas with the ~130 meV hard gap ($N$=20, colored purple) are always observed away from apical oxygen defects while the V-shaped, metallic LDOS ($N$=1, also the red regions) are centered around the defect sites. Upon further examination of the spectra, we identify four characteristic energy scales ($E_{peak}$, $E_{kink}$, $E^+$ and $E^-$) indicated by arrows in Fig. 5f and g. Plotting these as a function of $N$ reveals how these energy scales evolve with the crossover from an insulator to a metal (Fig.5h). Focusing on the spectral evolution of V-shaped metallic LDOS ($N$=1) from underlying parent LDOS ($N$=20), we find that the spectral weight transfer occurs continuously over a wide energy region, as schematically shown in Fig. 5i. From this data it is clear that the LDOS evolution (Fig. 5f-h) cannot be understood in terms of a simple filling change within a rigid band picture.

A qualitative insight into the trends in $E_{peak}$ and $E_{kink}$ can be obtained by our GGA and slab calculations. As shown in the GGA calculations in Fig. 3c, both the gap size (labeled $E_{kink}^{calc}$) and the peak energy (labeled $E_{peak}^{calc}$) shift towards $E_F$ with decreasing $U$, mimicking the experimentally observed LDOS evolution (Fig. 5f-h). However, we note that since $U$ is treated in a mean field level in GGA, the main effect of correlations in GGA is to renormalize the crystal field splitting in an orbital dependent manner [30]. Thus, in addition to charge doping, defect-induced changes in the crystal fields may be indirectly reflected by varying $U$ in the GGA calculations (see Supplementary Information for a detailed discussion). Preliminary slab calculations indicate that in the absence of the apical oxygen, the Ir atoms shift toward the missing oxygen site and metallic bands are formed at the Ir atom below the vacancy due to a downshift of the conduction bands (see Supplementary Information Fig S.7). The calculations thus support our observations that the low energy density of states in Ir327 is highly sensitive to small changes in local parameters brought about by the distortion created by vacancies. The sensitivity of the band structure indicates the tunable nature of the 327 compounds, an important aspect in realizing the novel electronic phases emergent at the boundary of a destabilized spin-orbit interaction driven correlated insulator.

On a final note, we comment on the energy scales $E^+$ and $E$ which emerge as the system gets more metallic and get progressively more symmetric about the Fermi energy. These features are not seen in the GGA calculations. While the origin of this intriguing particle–hole symmetric gap like feature is far from clear, this complex behavior with multiple emergent energy scales is unlike the disorder induced bound states picture in a simple band insulator. These trends are instead reminiscent of observations in correlated 3d transition metal oxides[14,31,32] where the LDOS evolves quite dramatically from the parent Mott insulating state due to the complex many-body effects associated with strong correlations.

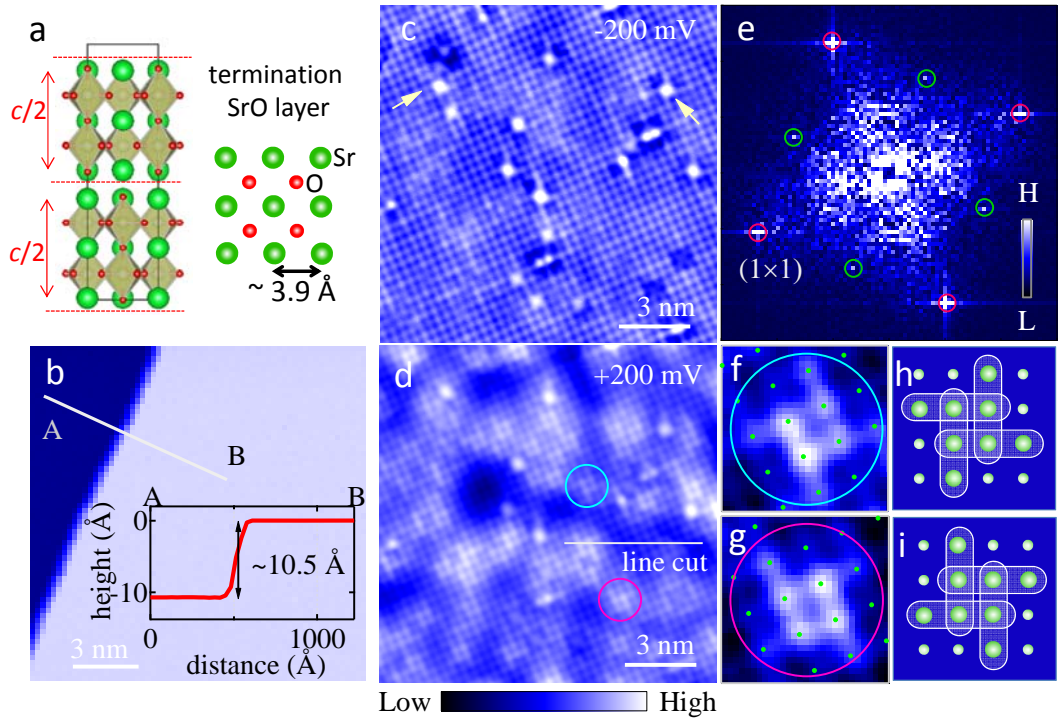

**Figure 1 | Topographic images of Sr$_3$Ir$_2$O$_7$. (a)** Crystal structure of Sr$_3$Ir$_2$O$_7$ **(b)** Typical step edge in topography, whose step height is half of the unit cell as indicated by arrows in (**a**). **(c)** Topographic image at bias voltage $V_B$=-200mV. **(d)** Topography at $V_B$=+200mV in the same area as **(c)**. The two typical types of chemical disorder are indicated by circles and arrows in **(c)** and **(d)**. **(e)** Fourier transform of **(c)**, showing Bragg peaks representing 1x1 (pink circles) and √2x√2 (blue circles) order. **(f)** and **(g)** Square shaped defects with two different chiralities. Topographic lattice sites are indicated by dots. **(h)** and **(i)** Schematic representation of chiral defects shown in **(f) and (g).** The green circles represent the atoms seen in the termination SrO layer.

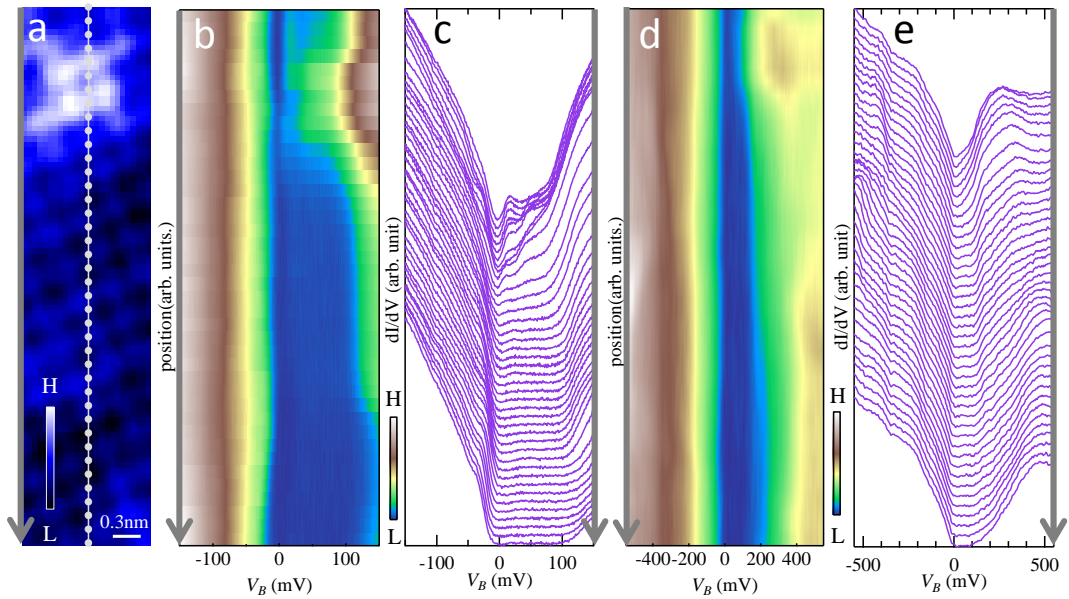

**Figure 2 | Tunneling spectra across chemical defect.** (**a**) A section of the topographic image shown in **Fig. 1d** (bias voltage $V_B$=+200 mV) where the linecuts shown in **(c)** and **(e)** were obtained. **(b), (d)** Image of the conductance intensity along the line. **(c)** Tunneling spectra along the line shown in **(a)** over a narrow energy region with energy resolution ~1 meV. **(e)** Tunneling spectra along the line over a and wide energy region (with ~10 meV resolution)

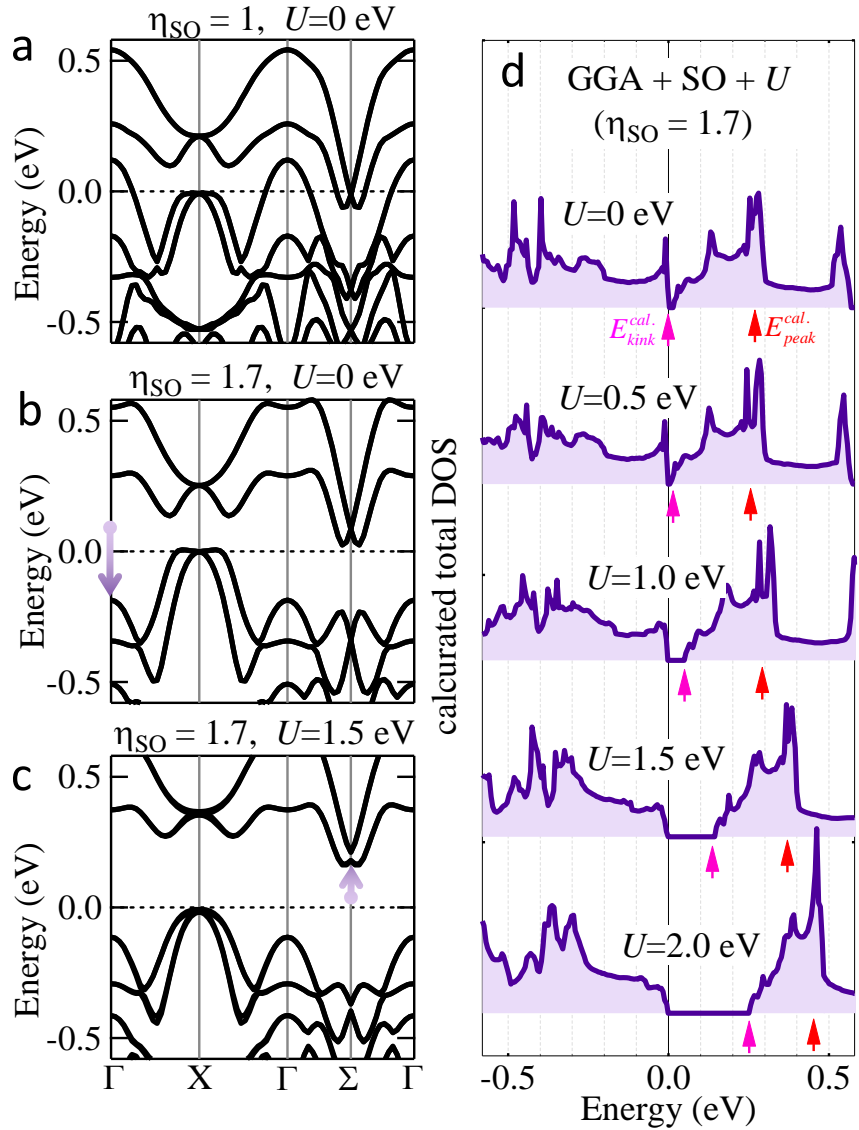

**Figure 3 | LDA band calculation along high symmetry lines**. **(a)** Band structure with SO coupling obtained self consistently. **(b)** Band structure with enhanced SO coupling. Here, SO=1.7 represents a SO strength which is 1.7 times the GGA self-consistently obtained value. **(c)** Band structure with on-site energy $U$=1.5(eV) added to (b). **(d)** Local density of states calculations for different values of coulomb U. The pink and red arrows refer to $E_{kink}^{calc}$ and $E_{peak}^{calc}$ respectively.

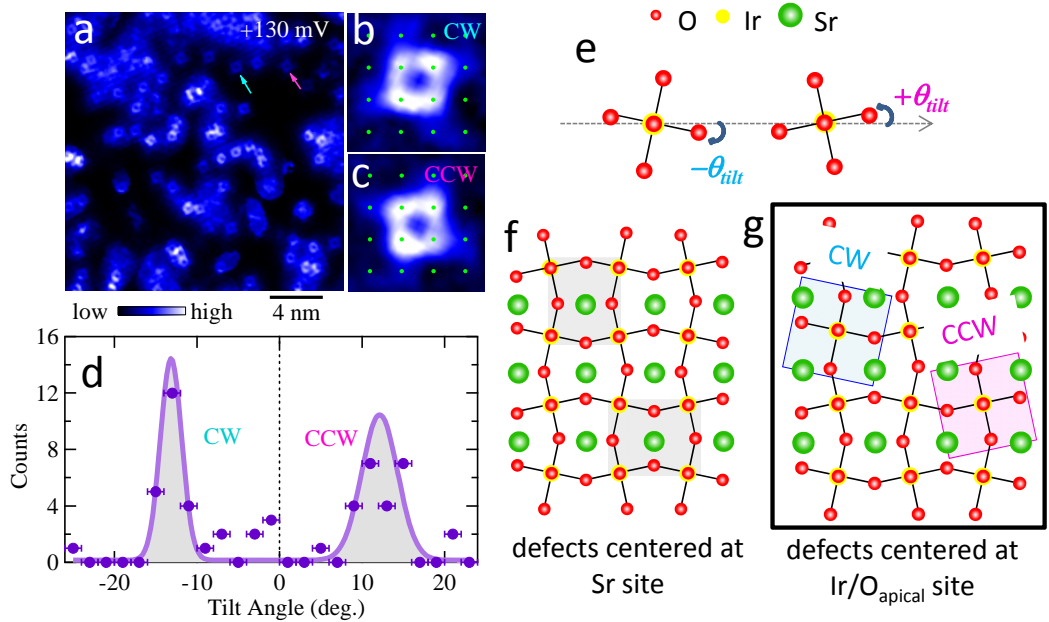

**Figure 4 | Visualization of the tilt angle of the underlying iridium oxide layer via crystal defects.** (**a**) d$I$/d$V$ map at +130 mV showing the local density of states signatures of the chiral defects shown in Fig. 1. Note that this area is different from **Fig. 1c and d**. (**b**), (**c**) d$I$/d$V$ maps of two defects tilted clockwise (CW) and counter clockwise (CCW) from in-plane Sr-O bond direction. The dots are the locations of the atoms from the simultaneously recorded topography. (**d**) Histogram of the tilt angles obtained from impurities in (**a**)(see main body and supplementary material for more details.) (**e**) Schematic drawing representing characteristic tilt angle $\pm\theta_{tilt}$, which is the in-plane O-Ir-O bond angle. (**f**), (**g**) Schematic top view of SrO layer and the IrO layer. The expected local symmetry of the two inequivalent sites centered on Sr (**f**) and centered on the apical oxygen site, (**g**).

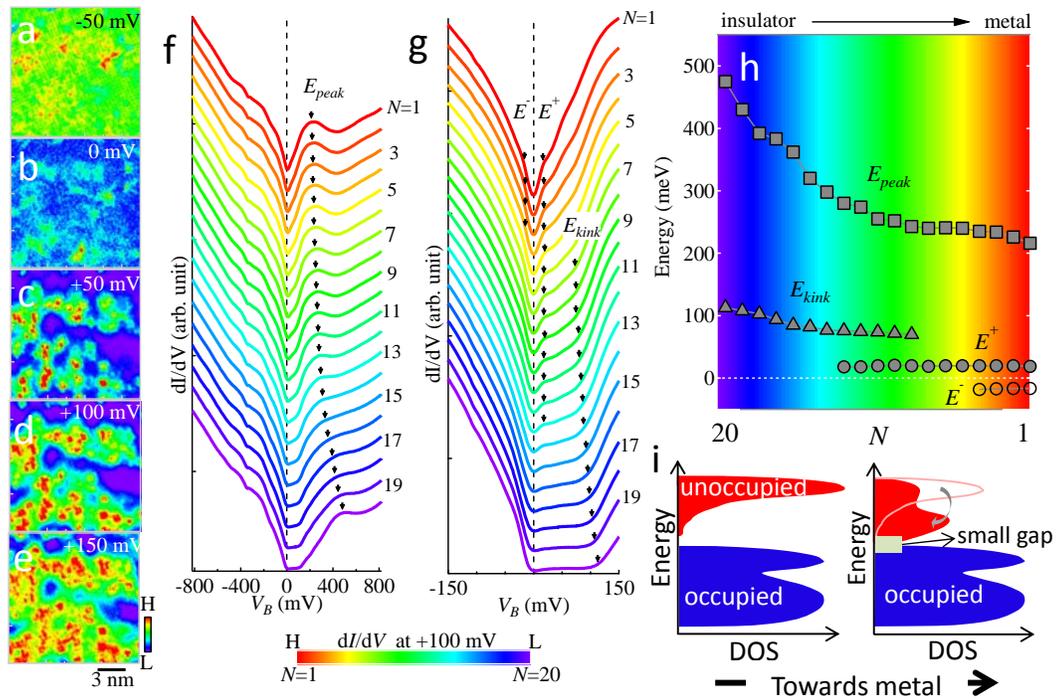

**Figure 5 | Spatial evolution of tunneling spectra (a)-(e)** Energy dependence of conductance map obtained in the same area shown in **Fig. 1c and d**. (**f**), (**g**) Averaged spectra classified by similar values of conductance at +100 mV. Here, spectra are split into 20 groups with equal population, from highest (*N*=1, red) to lowest (*N*=20, blue) conductance. (**f**) Narrow energy spectra with higher energy resolution (~1 meV). (**g**) Wide energy spectra (with ~10meV resolution). (**h**) Evolution of the characteristic energy scales ($E_{peak}$, $E^+$, $E^-$, and $E_{kink}$) represented by arrows in (**f**) and (**g**) shown as a function of N. (**i**) Schematic spectral evolution based on experimental data.

# Supplementary Information

## S1. Supplementary Methods

## S2. Supplementary Discussion

## Supplementary Figures

# S1. Supplementary Methods
## S1-1. Experimental details

Typical topographic images of sample surfaces cleaved at room temperature (**Fig. S1a and b**) and liquid nitrogen temperatures (**Fig. S1c**) are shown. Based on our experience, cold cleaving is necessary to obtain clean surfaces. In the RT cleaved samples, Sr or O atoms often form one dimmentional patterns. These chains are quite densly packed and prevent us from clearly reolving the atoms in the topographic images. On the other hand, the surface if the cold cleaved samples is far cleaner and the atoms are clearly resolved. All the data shown in the main paper were therefore obtained from samples cleaved at liquid nitrogen temperatures.

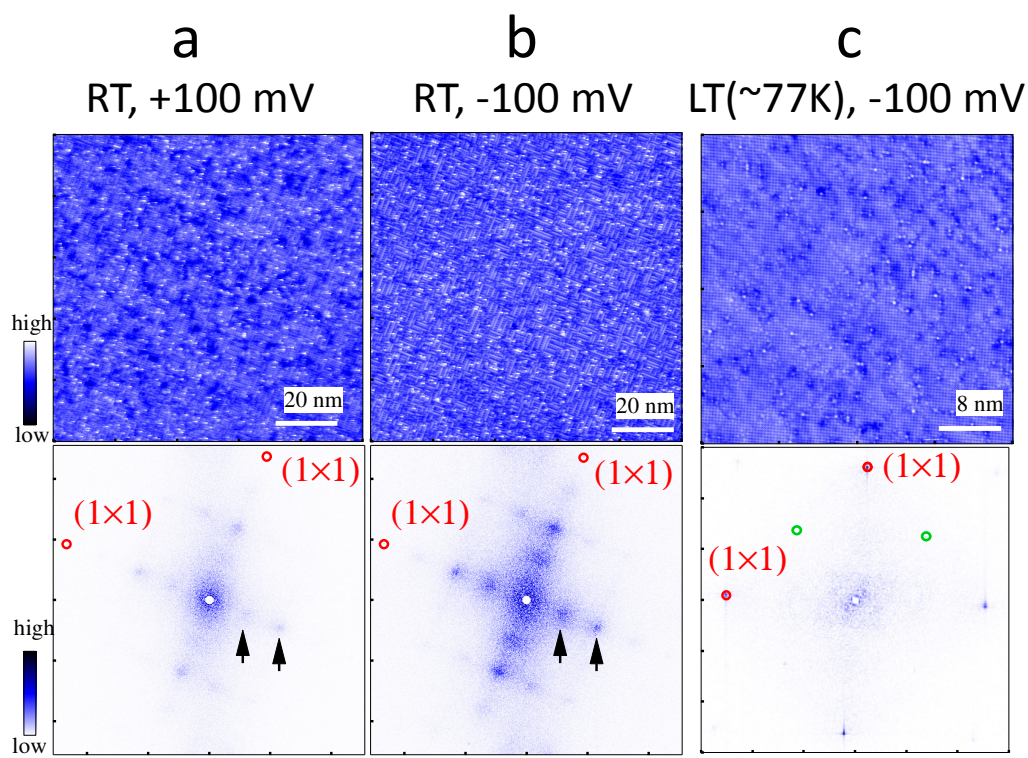

**Figure S1 | Typical topographic images for samples cleaved at room temperature (RT) and liquid nitrogen temperature (LT).**
Upper row shows topographic images and lower row shows Fourier transform (FT) images. Topography of sample cleaved at RT with bias voltage + 100 mV **(a)** and -100 mV **(b)**. **(a)** and **(b)** were obtained in the same area. **(c)** Topographic image of the surface cleaved at LT with bias voltage - 100 mV. Arrows in the lower panels of (a) and (b), refer to broad periodic structures at 1/4 and 1/2 the reciprocal lattice periodicity which are absent in the LT cleaved surfaces.

## S1-2. Visualization of rotated oxygen octahedra: angle measurement

The large area d$I$/d$V$ maps were first corrected for non-linear image distortion following **[S1]**; at the same time a simple linear transformation was applied to guarantee a square lattice. After these two correction procedures we used the following process to obtain the rotation angles of the impurities. We constructed a reference square shape of approximately the same size as the impurity dI/dV image. We then obtained a cross correlation coefficient between the reference image **(Fig. S2c)** and the local dI/dV map around each defect **(Fig. S2b)** as a function of the former's rotation angle. The tilt angle was defined as the angle that gives the maximum correlation coefficient **(Fig. S2d)**. Small details of this reference shape did not significantly affect the obtained tilt angle. We performed this procedure for the clearly resolved chiral defects shown by circles in **Fig. S2a**.

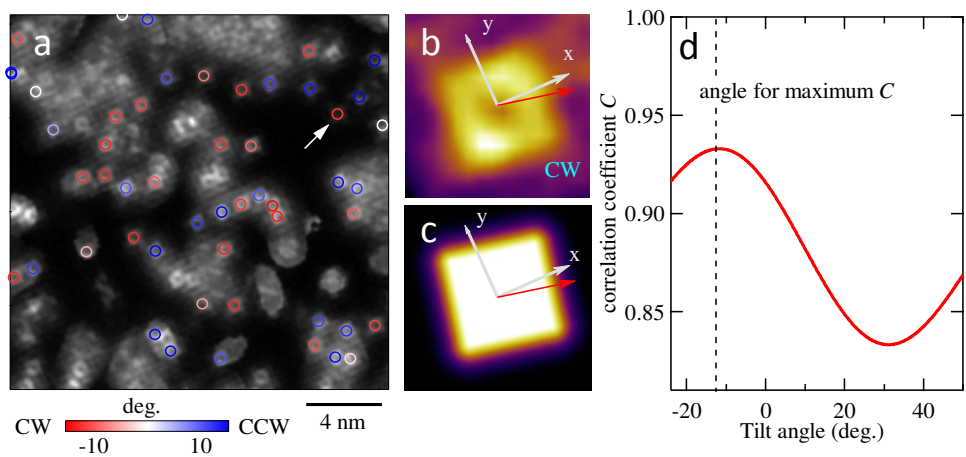

**Figure S2 | Method for determining tilt angle via defects images.**
**(a)** Conductance image, which is same as **Fig. 4(a)**, but with defect positions (circles). The color of the circle indicates the measured tilt angle. **(b)** One of the defects, indicated by the arrow in **(a)**. **(c)** The reference square, rotated through the maximum-correlation angle. **(d)** Correlation coefficients between **(b)** and **(c)** as a function of rotation angle; the angle is measured with respect to the lattice direction.

## S1-3. Categorizing spatial evolution of LDOS via the conductance at +100mV

As shown in **Fig. S3a and b**, large spectral spatial variations appear between $E_f$ and ~+400 mV. To obtain an overall picture of the evolution of the spectral lineshape and to correlate this evolution with the topography, it was necessary to categorize the tunneling conductance (dI/dV) spectra into similar groups based on the spectral shape. The best method we found was to use the dI/dV value at +100 mV to sort the spectra. We split the spectra into 20 groups (with indexing $n$=1 to 20) as shown in **Fig. 5 (also in Fig. S4)**. To show the validity of this sorting method, we show spectral intensity histogram images of selected groups in **Fig. S4**. As shown in **Fig. S4b-d,** the spectra with the same value of $n$ show only small variations in their lineshape. This was confirmed for all the 20 categories. The spectra in each category of n were averaged and plotted in Fig. 5 of the main text where a comparison with the topography shows the clear correlation between the lineshape and the presence of impurities.

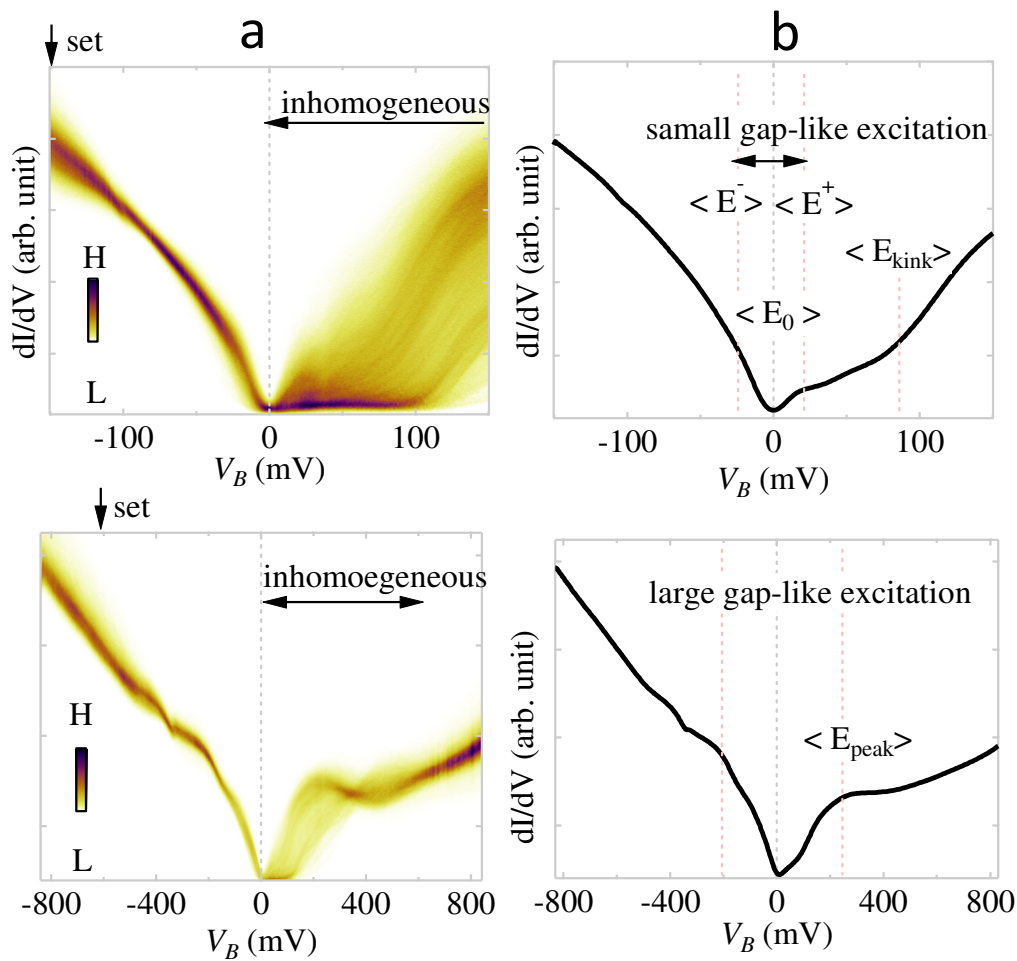

**Figure S3 | Spectral variation and averaged spectra.**
(a) is histogram of the 256*256 tunneling spectra obtained in the same area in **Fig. 1, 2, 3, and 5**. (b) is spatially averaged spectra of (a). Characteristic energy scales are shown in (b). Upper row is for narrow but higher energy resolution spectra. Lower row is for wide but lower energy resolution spectra.

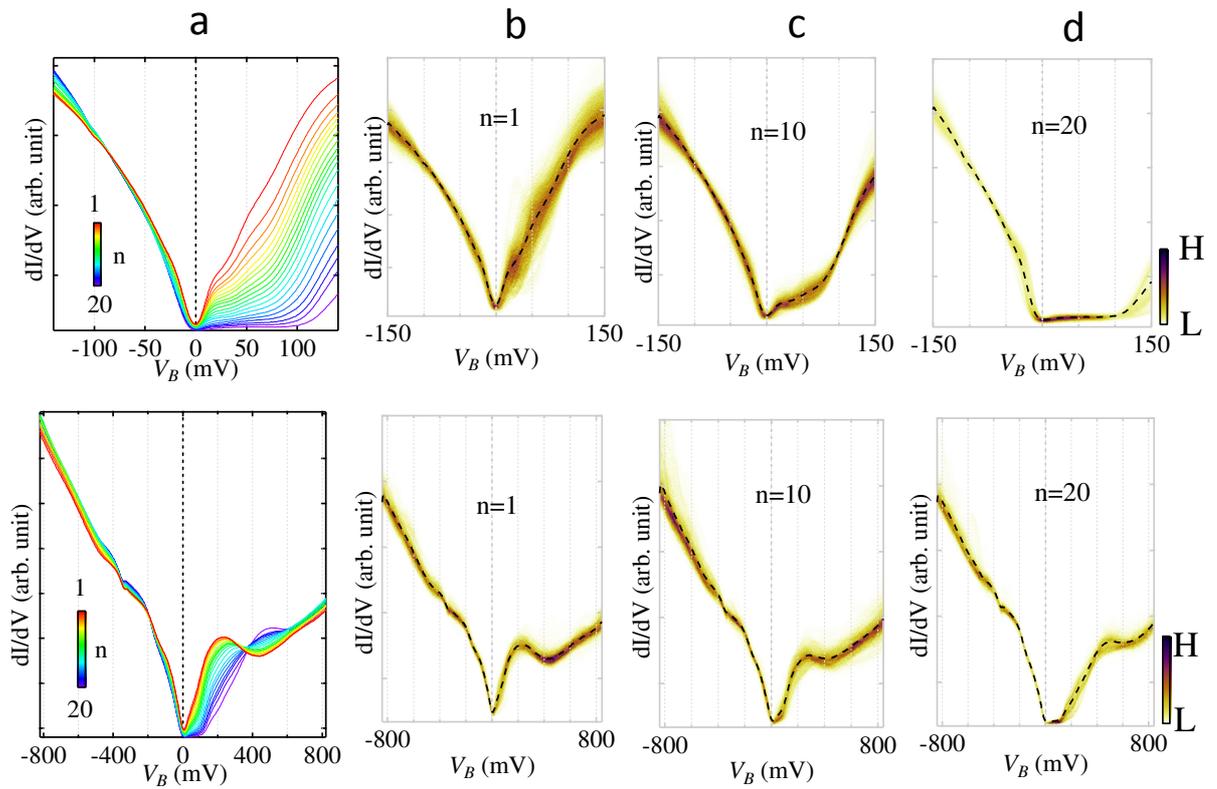

**Figure S4 | Categorization of the spectra and its validity.**
**(a)** Average spectra of the 20 groups shown in **Fig. 5**. Histogram of the spectra within the same group are shown for $n$=1**(b)**, 10**(c)**, and 20**(d)**. Upper row is for narrow but higher energy resolution spectra. Lower row is for wide but lower energy resolution spectra.

**S1-4. Current dependence of the tunneling spectral shape**
Since the sample is not a good electric conductor, tip-induced band bending (TIBB) may occur and distort the spectral shape. The quantitative evaluation of TIBB on an unknown sample is quite difficult experimentally. However, we can qualitatively estimate the degree of TIBB effects by checking tip-sample distance dependence of the spectral shape**[S2]**. Here, tip-sample distance can be modified by changing set point current and/or bias voltage. We did not see any significant current dependence with the tips and samples used in this study (**Fig. S5 b and c**).

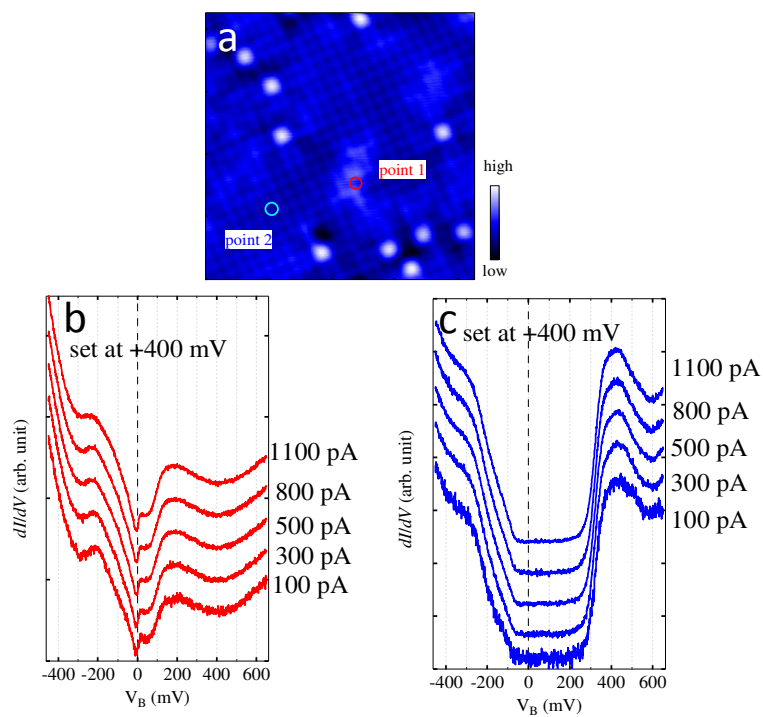

**Figure S5 | Current dependence of the spectra.**
**(b)** and **(c)** shows the current dependence of the tunneling spectra obtained at two different points shown in **(a)**.

# S2. Supplementary Discussion

**First principles calculation**
**S2-1. Persistent metallic state without increasing Hubbard *U***
In this study, SO coupling is included in the self-consistent cycles of the first-principles band calculation. Our calculations allow us to vary the strength of SO by adding a prefactor $\eta_{SO}$. For example, $\eta_{SO}$ =2 means the strength of SO is doubled. GGA results have suggested that the band structure remains metallic unless the Hubbard *U* (from *U*=0) and $\eta_{SO}$=1 are increased. Therefore, it is necessary to invoke increased both an *U* (>0) and/or $\eta_{SO}$ (>1) to explain experimentally observed intrinsic LDOS (away from chemical disorder). Below we address how U and $\eta_{SO}$ was varied in this study to explain experimental data.

**S2-2. Qualitative estimation of $\eta_{SO}$ and U.**
As shown in **Fig. 3a** and **b**, the SO coupling largely determines the bands around the Γ point. On the other hand, *U* mainly affects the bands around the Σ point. To estimate both $\eta_{SO}$ and *U*, we considered several constraints. First constraint is our observation of ~120mV hard gap in the intrinsic LDOS (away from disorder and corresponding to n=20 in **Fig. 5f and g**). With $\eta_{SO}$=1, the indirect gap is always found to be negligible because of a band overlapping the Fermi energy at the Γ point. Furthermore, since U does not alter bands at the Γ point this negligible band gap was found to persist with increasing *U* (to within limits of realistic values). To creat a gap in the LDOS, we had to increase $\eta_{SO}$ from $\eta_{SO}$=1, a possibility also suggested by recent angle resolved photoemission spetroscopy (ARPES) measurements on Ir327 **[ref_ARPES]**. To match ARPES data and semi-quantitatively explain the hole-like band edge position at the Γ point, $\eta_{SO}$=1.7 was found to be the best value (**Fig. 3b**).

However, the band with $\eta_{SO}$=1.7 and *U*=0 still does not capture the ~120 mV hard gap due to the electron-like unoccupied state band at the Σ point as seen in **Fig. 3b**. It is Hubbard *U* that mostly governs the band at the Σ point. We therefore explore the theoritical LDOS with a fixed $\eta_{SO}$=1.7 and a changing *U* as shown in **Fig. 3d** and find that the best value of U is between 1.5eV -2.0eV.

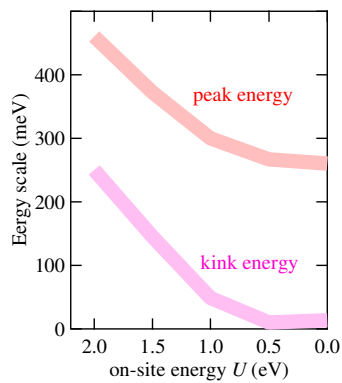

**Figure S6** | **Characteristic energy scales obtained from simulated LDOS with changing U.** Peak and kink energy is obtained from **Fig. 3d**, where these two energy scales are indicated by arrows.

Another important parameter is octahedral tilt, which creates √2x√2 supersutructure. Note that all the calculations have been done assuming 12 ° rotation, which is from X-ray data in the literature**[ref_X-Ray]**. As shown in **Fig. 4d,** this value is quite consistent with our STM experiment. Moreover, within the possible realistic tilt angle variation on the surface (within ±3°), we checked that tilt angle has a minor role for band alternation compared with $\eta_{SO}$ and $U$.

A possible qualitative explanation how apical oxygen defects make the LDOS more metallic is obtained from 2x2x2 super-cell slab calculation with the inclusion of an apical oxygen vacancy (**Fig. S7**). After putting apical oxygen vacancy (upper left of **Fig. S7a** and **b**), the Ir atom just under the apical oxygen vacancy moves towards apical oxygen vacancy center (see arrow in **Fig. S7a**). Together with this atomic deformation, metallic bands are found at the Ir atom below the vacancy due to the downshift of the conduction band at M point (**Fig. S7c** and **d**). The metallic LDOS arisen by apical oxygen defects are schematically drawn in **Fig. S7e and f.**

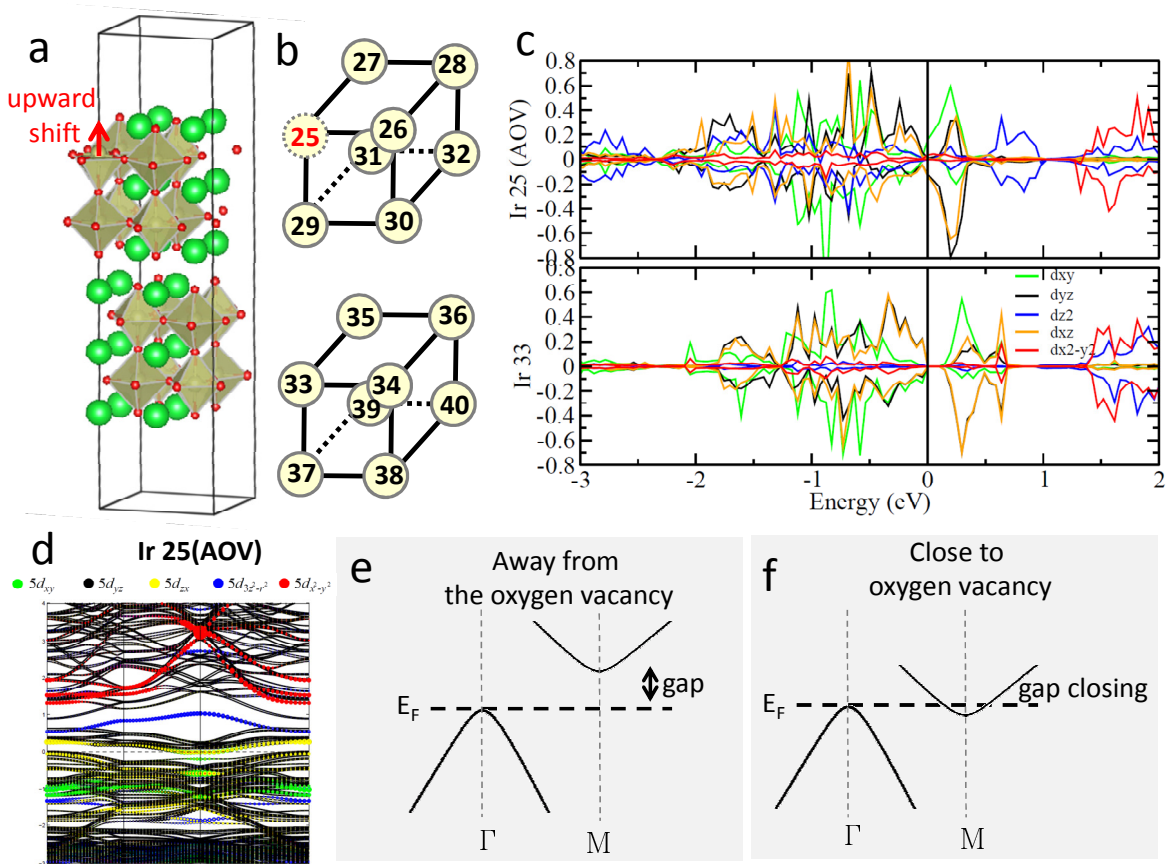

**Figure S7 | 2x2x2 super-cell slab calculation with an apical oxygen vacancy. (a) (b)** Model structure with an apical oxygen defects at site 25 (upper left). **(c)** Partial LDOS at 25 (upper) and 33 (lower) at the Ir sites. **(d)** Orbital dependent band structure at Ir site just under the apical oxygen vacancy. Schematic picture of the band structure away from apical oxygen vacancy **(e)** and close to apical oxygen vacancy **(f)**. Here, this calculation is done with U=1.5 eV and $\eta_{SO}$ =1.7 with in-plane 12° octahedral tilt.

### S2-3. Small energy scale at Fermi energy

As shown in **Fig. 5**, we fiind that an almost particle-hole symmetric small gap starts to appear at a few tens of meV around the Fermi energy. The reproducibility of this finding was checked with many different tips and samples. Preliminary data indicate that this small gap-like anomaly does not show an obvious sensitivity to an external magnetic field pararell to c-axis up to 7T. While the origin of this feature is not yet clear, this structure hints at the possibilty of an emergent phase in a regime with with more vacancies where the metallic regions doimnate the sample.

**[S1] drift correction**
"Intra-unit-cell electronic nematicity of the high-Tc copper-oxide pseudogap states"
M. J. Lawler, K. Fujita, J. Lee, A. R. Schmidt, Y. Kohsaka, C. K. Kim, H. Eisaki, S. Uchida, J. C. Davis, J. P. Sethna, and Eun-Ah Kim
Nature 466,347–351(15 July 2010)
http://www.nature.com/nature/journal/v466/n7304/full/nature09169.html
**[S2] tip induced band bending**
for example, see
"Observation of spatially inhomogeneous electronic structure of Si(100) using scanning tunneling spectroscopy"
K. Nagaoka, M. J. Comstock, A. Hammack, and M. F. Crommie
Phys. Rev. B 71, 121304 (2005)
http://prb.aps.org/abstract/PRB/v71/i12/e121304